\documentclass[sigconf,nonacm]{acmart}
\AtBeginDocument{%
  }

\setcopyright{acmlicensed}
\copyrightyear{2018}
\acmYear{2018}
\acmDOI{XXXXXXX.XXXXXXX}
\acmConference[Conference acronym 'XX]{Make sure to enter the correct
  conference title from your rights confirmation email}{June 03--05,
  2018}{Woodstock, NY}
\acmISBN{978-1-4503-XXXX-X/2018/06}

\usepackage{pifont}
\usepackage[table]{xcolor} % for \rowcolors in tables
\usepackage{pgfplots} % for axis and tikzpicture environments
\usepackage{enumitem}
\usepackage[edges]{forest}
\usepackage{graphicx}
\usepackage{booktabs} % For \toprule, \midrule, \botrule
\usepackage{threeparttable} % For table footnotes
\usepackage{algorithm}
\usepackage{algorithmicx}
\usepackage{amsmath}
\usepackage{algpseudocode}
 \usepackage[export]{adjustbox}
\usepackage{svg}
\usepackage{diagbox}

\usepackage{tikz}
\usetikzlibrary{trees}
\usetikzlibrary{shadows}
\usepackage[edges]{forest}
\definecolor{crq}{HTML}{ededed}
\definecolor{crq2}{HTML}{dee2e6}
\definecolor{crq4}{HTML}{425168}
\definecolor{crq3}{HTML}{535662}
\definecolor{g_color}{HTML}{2a918c}
\definecolor{r_color}{HTML}{FF4242}
\definecolor{solicitationcolor}{HTML}{019fe2}

\definecolor{darkgray}{RGB}{70,70,70}
\definecolor{mediumgray}{RGB}{130,130,130}
\definecolor{lightgray}{RGB}{200,200,200}
\definecolor{verylightgray}{RGB}{240,240,240}
\definecolor{accentgray}{RGB}{100,100,100}

\definecolor{crq}{HTML}{ededed}
\definecolor{crq2}{HTML}{dee2e6}
\definecolor{crq4}{HTML}{425168}
\definecolor{crq3}{HTML}{535662}
\definecolor{rolecolor}{HTML}{2a918c}
\definecolor{creationcolor}{HTML}{FF4242}
\definecolor{solicitationcolor}{HTML}{019fe2}
\usepackage{float}

\usetikzlibrary{shapes,shadows,arrows}
\tikzstyle{rqbox} = [draw=crq2, fill=crq, very thick,
    rectangle, rounded corners, inner sep=10pt, inner ysep=12pt]
\tikzstyle{titlerq} =[fill=crq2, draw=crq2,  rounded corners, inner sep=4pt]

 \setcopyright{none} 
 
\setcopyright{none}
\usepackage{adjustbox}
\usepackage{tcolorbox}
\usepackage{subcaption}
\usepackage{siunitx}
\usepackage{enumitem}
\usepackage{makecell}
\usepackage[nameinlink]{cleveref}
\usepackage{multirow}
\usepackage{threeparttable}
\usepackage{xcolor} 
\usepackage{amsmath}
\usepackage{xurl}
\usepackage{colortbl}
\usepackage[textsize=1in]{todonotes}
\usepackage{graphicx} 
\usepackage{hyperref} 
\usepackage{array}    
\usepackage{pifont}
\usepackage{float}
\usepackage{amsmath}
\usepackage{fancyvrb}
\usepackage[utf8]{inputenc}
\usepackage{listings}
\usepackage{tabularx}
\usepackage{xcolor}

\definecolor{bleu}{rgb}{0.2,0.2,0.7}
\lstdefinestyle{mypython}{
  language=Python,
  basicstyle=\ttfamily\small,
  keywordstyle=\color{blue},
  showstringspaces=false,
  breaklines=true,
  frame=none
}

\newboolean{showcomments}
\setboolean{showcomments}{true}
\ifthenelse{\boolean{showcomments}}
 { \newcommand{\mynote}[2]{
      \fbox{\bfseries\sffamily\scriptsize#1}
        {\small$\blacktriangleright$\textsf{\emph{#2}}$\blacktriangleleft$}}}
        { \newcommand{\mynote}[2]{}}

\definecolor{highcolor}{RGB}{204,255,204} % Light green for highest values
\definecolor{lowcolor}{RGB}{255,204,204} % Light red for lowest values
\definecolor{codegreen}{rgb}{0,0.6,0}
\definecolor{codegray}{rgb}{0.5,0.5,0.5}
\definecolor{codepurple}{rgb}{0.58,0,0.82}
\definecolor{backcolour}{rgb}{0.95,0.95,0.92}
\begin{document}

%\title{Security risks in the software build process}
\title{Characterizing Build Compromises Through Vulnerability Disclosure Analysis}

\author{Maimouna Tamah Diao}
\authornote{Both authors contributed equally to this research.}
%\author{Moustapha Awwalou Diouf}
%\email{moustapha.diouf@uni.lu}
\affiliation{%
  \institution{University of Luxembourg}
  \country{Luxembourg}
}
\email{maimouna.diao@uni.lu}

\author{Moustapha Awwalou Diouf}
\authornotemark[1]
\affiliation{%
  \institution{University of Luxembourg}
  %\city{Hekla}
  \country{Luxembourg}}
\email{moustapha.diouf@uni.lu}

\author{Iyiola Emmanuel Olatunji}
\affiliation{%
   \institution{University of Luxembourg}
   \country{Luxembourg}
}
\email{emmanuel.olatunji@uni.lu}

\author{Abdoul Kader Kaboré}
\affiliation{%
 \institution{University of Luxembourg}
 %\city{Doimukh}
 %\state{Arunachal Pradesh}
 \country{Luxembourg}}
 \email{abdoulkader.kabore@uni.lu}

\author{Gervais Mendy}
\affiliation{%
  \institution{University Cheikh Anta Diop}
  %\city{Haidian Qu}
  %\state{Beijing Shi}
  \country{Senegal}}
  \email{gervais.mendy@ucad.edu.sn}

\author{Jacques Klein}
\affiliation{%
  \institution{University of Luxembourg}
  %\city{San Antonio}
  %\state{Texas}
  \country{Luxembourg}}
\email{jacques.klein@uni.lu}

\author{Tegawendé F. Bissyandé}
\affiliation{%
  \institution{University of Luxembourg}
  %\city{Hekla}
  \country{Luxembourg}}
\email{tegawende.bissyande@uni.lu}

\renewcommand{\shortauthors}{Diao et al.}

\begin{abstract}
The software build process transforms source code into deployable artifacts, representing a critical yet vulnerable stage in software development. Build infrastructure security poses unique challenges: the complexity of multi-component systems (source code, dependencies, build tools), the difficulty of detecting intrusions during compilation, and prevalent build non-determinism that masks malicious modifications. Despite these risks, the security community lacks a systematic understanding of build-specific attack vectors, hindering effective defense design.

This paper presents an empirically-derived taxonomy of attack vectors targeting the build process, constructed through a large-scale CVE mining (of 621 vulnerability disclosures from the NVD database). We categorize attack vectors by their injection points across the build pipeline, from source code manipulation to compiler compromise. To validate our taxonomy, we analyzed 168 documented software supply chain attacks, identifying 40 incidents specifically targeting build phases. Our analysis reveals that 23.8\% of supply chain attacks exploit build vulnerabilities, with dependency confusion and build script injection representing the most prevalent vectors.

Dataset available at: \url{https://anonymous.4open.science/r/Taxonomizing-Build-Attacks-8BB0}.
\end{abstract}


\received{20 February 2007}
\received[revised]{12 March 2009}
\received[accepted]{5 June 2009}

\maketitle

\section{Introduction}
The software build process—transforming source code into distributable artifacts—represents a critical attack surface in modern software development. This process orchestrates multiple components, including source code, third-party dependencies, build tools, and execution environments, each introducing distinct vulnerability classes. In contemporary CI/CD pipelines, where automation drives deployment velocity, these vulnerabilities become particularly concerning: malicious scripts execute with elevated privileges, sensitive credentials persist in build configurations, and intermediate artifacts undergo manipulation—all potentially compromising the final product without modifying the source code itself.

The security implications of build vulnerabilities are amplified by a fundamental challenge: build non-reproducibility. Recent studies demonstrate that 15-30\% of software packages cannot be reliably reproduced~\cite{10.1145/3180155.3180224,10.1145/3510003.3510102}, while the Debian Reproducible Builds project reveals that even with dedicated effort, only 95\% of packages achieve reproducibility~\cite{dl1,dl2}. This non-determinism creates an ideal environment for stealthy attacks, as legitimate variations in build outputs mask malicious modifications. The problem intensifies in CI/CD environments where rapid iteration cycles and complex tool chains reduce visibility into individual build steps.

Despite these critical risks, the security research community has predominantly focused on dependency-related attacks~\cite{ladisa2023sok,neupane2023beyond}, leaving build infrastructure vulnerabilities systematically unexplored. While recent work has proposed taxonomies for supply chain attacks targeting package managers and dependency confusion, no comprehensive analysis exists for attack vectors specific to the build process itself—particularly in automated CI/CD contexts where 75\% of organizations now deploy code~\cite{software2025}.

To address this gap, we conducted a large-scale mining study of Common Vulnerabilities and Exposures (CVEs) to systematically identify and categorize attack vectors targeting build infrastructure. We analyzed 621 CVEs from the National Vulnerability Database (2015-2024) related to build systems, CI/CD platforms, and compilation tools. Our mining methodology combines automated filtering using build-specific keywords with manual analysis to extract attack patterns and exploitation techniques. This CVE-derived knowledge is then validated against 40 real-world build attacks identified from a comprehensive dataset of 168 software supply chain incidents~\cite{atlantic_council_dataset}.

Our dual approach—mining theoretical vulnerabilities from CVEs and mapping them to observed attacks—reveals that build-phase compromises account for 23.8\% of supply chain attacks, yet receive disproportionately little attention in security research. The attacks we identified exploit diverse vectors: from GitHub Actions workflow poisoning to compiler backdoors, from Docker image tampering to build cache poisoning. These findings underscore the urgent need for systematic understanding of build security.

To develop a comprehensive and reliable taxonomy of build attack vectors, we structured our investigation around three complementary research questions that guide the taxonomy's construction, refinement, and validation:

\begin{itemize}[leftmargin=*]
    \item {\bf RQ1:} \textit{What are the distinct attack vector patterns that characterize build infrastructure vulnerabilities?} Through systematic CVE mining, we identify and categorize the main vulnerability classes that enable build-phase compromises. This analysis establishes the structural basis for our taxonomy.
    \item \textbf{RQ2:} \textit{How do build attack vectors manifest and propagate through modern CI/CD pipelines?} We analyze attack prevalence, entry points, and propagation patterns to understand which build components are most frequently exploited and why. These insights refine our taxonomy by revealing the practical relationships between attack vectors and their impact (information theft or artifact compromise). 
    \item \textbf{RQ3:} \textit{To what extent can our CVE-derived taxonomy capture the full spectrum of real-world build attacks?} We assess the completeness and applicability of our taxonomy by examining its coverage of documented incidents. This validation step ensures that our taxonomy is not merely theoretical but accurately represents the threat landscape practitioners face.
\end{itemize}

Our contributions include:

\begin{enumerate}
    \item A systematic CVE mining methodology extracting 621 build-related vulnerabilities into a structured taxonomy.
    \item Empirical validation demonstrating 92\% coverage of real-world attacks through our taxonomy.
    \item A curated, publicly available dataset linking CVEs to build attacks, enabling reproducible research.
    \item Actionable insights for prioritizing security controls in CI/CD pipelines based on attack prevalence.
\end{enumerate}

\section{Motivation}
The 2020 SolarWinds compromise demonstrates the stealthy reality of build-phase attacks and their catastrophic potential. Attackers injected the SUNBURST backdoor directly into the build process through dynamic code generation, bypassing source code repositories entirely~\cite{SUNBURST_backdoor}. The malicious code existed only during compilation—inserted via a temporary file that was deleted post-build—leaving no trace in version control systems. This technique produced legitimately signed binaries that infected over 18,000 organizations across government, technology, and financial sectors~\cite{9382367}.

This attack exemplifies a critical blindspot in current security models: build processes are treated as deterministic transformations rather than complex systems with their own threat model. Security controls typically focus on repository integrity (pre-build) and artifact scanning (post-build), while the build phase itself—with its tool chains, environment configurations, and transient states—remains largely unexamined. The SolarWinds attackers exploited precisely this gap, operating in the ephemeral space between source and artifact where traditional security boundaries dissolve.

The absence of structured knowledge about build-specific attack vectors has practical consequences. Security teams lack frameworks for threat modeling build infrastructure, developers receive no systematic training on build security risks, and tooling vendors implement controls without understanding the full attack landscape. While individual organizations may develop ad-hoc protections based on past incidents, the community lacks a shared vocabulary and comprehensive understanding of the threat space.

A systematic taxonomy derived from empirical data addresses these gaps by providing: (1) a common reference framework for describing and analyzing build attacks, (2) structured knowledge for risk assessment and control prioritization, and (3) a foundation for developing targeted countermeasures. By mining CVEs and validating against real incidents, we transform fragmented vulnerability reports into actionable security intelligence.

\section{Methodology}
\begin{figure*}[!h]
 \begin{adjustbox}{width=\linewidth, center}
 \includegraphics{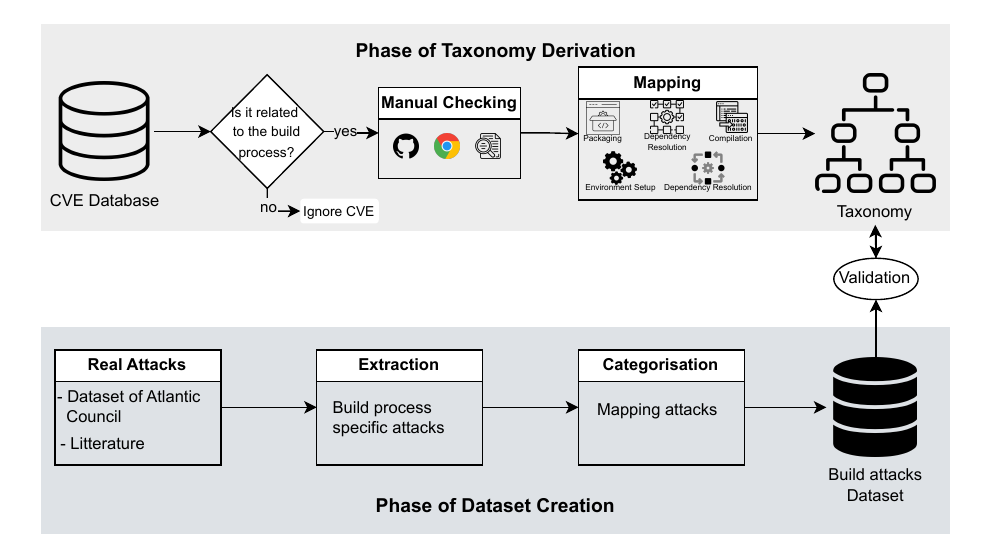}
\end{adjustbox}
\caption{\textbf{Two-phase approach for deriving and validating a taxonomy of build attack vectors. The upper workflow presents the taxonomy derivation from CVE analysis, while the lower workflow illustrates the validation dataset 
construction from real-world attacks.}}
%\tb{there are typos in the text = is it to build process?}}
\label{fig:build_step}
\end{figure*}

We followed a systematic mining approach to identify and categorize attack vectors specific to build processes, proceeding through two complementary mining efforts: (1) mining CVE repositories to construct the taxonomy, and (2) mining attack reports to build a validation dataset. Figure \ref{fig:build_step} illustrates this dual mining pipeline.

\subsection{Mining CVEs for Taxonomy Construction}

\subsubsection{CVE Extraction and Filtering}

We mined the National Vulnerability Database  using a targeted keyword-based extraction strategy. Our query set included build-specific terms such as ``maven'', ``gradle'', ``npm build'', ``bazel'', ``ci/cd'', and `compiler''. This initial extraction yielded 621 candidate CVEs requiring manual analysis.

The manual verification process proved particularly labor-intensive, requiring examination of CVE descriptions and metadata, official security advisories from vendors and maintainers, associated GitHub repositories and issue trackers where vulnerabilities were reported and discussed, and technical documentation including proof-of-concept exploits when available.

Each CVE underwent rigorous screening against four inclusion criteria:

\begin{itemize}[leftmargin=*]
    \item \textbf{Build Tool Relevance.} We retained CVEs that directly affect build tools or compilers, such as vulnerabilities in Maven's dependency resolution mechanism, Gradle's plugin system, or GCC's preprocessing directives. These vulnerabilities must originate from the build tool itself rather than from the code being built.
    
    \item \textbf{Build Stage Exploitation.} This criterion captures vulnerabilities that can be triggered during build execution, even if they originate elsewhere. For instance, a vulnerability in a logging library becomes \textit{build-relevant }if it can be exploited when the build system invokes logging functions during compilation or testing phases.
    
    \item \textbf{Build Configuration Targeting.} We included vulnerabilities that specifically exploit build scripts or configuration files, such as malicious code injection through Makefiles, CMakeLists.txt manipulation, or package.json script tampering. These attacks often leverage the elevated privileges and reduced scrutiny that build configurations typically receive.
    
    \item \textbf{Build Environment Impact.} This encompasses vulnerabilities that compromise CI/CD infrastructure components, including Jenkins pipeline exploits, GitHub Actions workflow poisoning, or Docker build context manipulation. Such vulnerabilities may not directly affect build tools but compromise the infrastructure orchestrating the build process.
\end{itemize}

This manual screening process, requiring approximately 400 person-hours, reduced our corpus to 33 confirmed build-related CVEs. The high rejection rate 94.68\% underscores the challenge of accurately identifying build-specific vulnerabilities from general CVE descriptions.

\subsubsection{Taxonomy Derivation Through Pattern Mining}

We employed a systematic pattern mining approach on the validated CVEs. For each CVE, we performed detailed analysis to extract the build stage where exploitation occurs (Environment Setup, Dependency Resolution, Compilation, Testing, Packaging), the attack vector characteristics and exploitation technique, and the impact on build process integrity.

This deep analysis phase required an additional 200 person-hours, as each CVE demanded careful examination to accurately map its attack pattern. We then applied hierarchical clustering to group CVEs exhibiting similar attack vectors into coherent categories. Through iterative refinement across five rounds of analysis, we organized these patterns into a structured taxonomy based on their exploitation point in the CI/CD pipeline.

The taxonomy construction employed open coding techniques, with two researchers independently categorizing CVEs before reconciling differences through discussion. Inter-rater reliability, measured using Cohen's Kappa, reached $75\%$, indicating substantial agreement.

\subsection{Validation Through Attack Report Mining}

\subsubsection{Validation Dataset Construction}

To validate our CVE-derived taxonomy against real-world threats, we constructed a ground truth dataset by mining software supply chain attack reports. This validation dataset, illustrated in the lower workflow of Figure \ref{fig:build_step}, required extensive manual analysis to identify build-specific compromises.

We mined the Atlantic Council's Cyber Statecraft Initiative database \cite{atlantic_council_dataset}, which documents 168 software supply chain incidents from 2015-2024. While comprehensive, this database does not explicitly categorize attacks by supply chain phase, necessitating manual classification.

For each incident, we conducted in-depth analysis involving technical report examination from multiple sources to corroborate attack details, attack timeline reconstruction to identify the exact point of compromise, build process involvement verification through forensic indicators, and exploitation technique identification from available artifacts and logs.

This labor-intensive mining process, requiring approximately 150 person-hours, involved disambiguating attack narratives and distinguishing build-phase compromises from other supply chain attacks. Only incidents with confirmed build process exploitation were retained.

Our analysis identified 40 attacks where the compromise specifically occurred during the build phase. We enriched this corpus with 2 recent cases documented in academic literature, yielding a final validation dataset of 55 build-specific attacks.

\subsubsection{Dataset Annotation and Enrichment}

Each attack in our validation dataset underwent meticulous annotation to document attack vector classification according to our taxonomy, affected build stage and components, exploitation technique, and entry point, and observable impact and persistence mechanisms.

%Two researchers independently annotated each incident, with conflicts resolved through consensus discussion. This annotation process required an additional \tb{xxx} person-hours to ensure accuracy and consistency.

\subsection{Validation Methodology}

We validated our taxonomy through systematic mapping of the ground truth dataset to our CVE-derived categories. This validation assessed:

% \tb{Try to use these criteria to the best effort possible to make it scientifically solid}
\begin{enumerate}[leftmargin=*]
    \item \textbf{Coverage:} We measured what percentage of real-world attacks could be accurately classified using our taxonomy categories. Each attack from our validation dataset was mapped to the taxonomy, with successful mappings indicating that our CVE-derived categories capture actual attack patterns observed in practice.
    
    \item \textbf{Completeness:} We identified whether any attack patterns present in the validation dataset were absent from our CVE-derived taxonomy. Attacks that could not be mapped to existing categories revealed potential gaps in CVE documentation or emerging attack vectors not yet captured in vulnerability databases.
    
    \item \textbf{Distribution:} We analyzed the prevalence of different attack vectors in real-world incidents versus their representation in CVE data. This comparison reveals which vulnerabilities are actively exploited in practice versus those that remain theoretical, informing risk prioritization for defenders.
\end{enumerate}

\subsection{Reproducibility}

The total effort for this study exceeded 400 person-hours of manual analysis, reflecting the inherent complexity of mining and categorizing build-specific security data. To support reproducibility and future research, we make available our complete list of 621 build-related CVEs with categorizations, the annotated validation dataset of 55 build attacks, our keyword lists and filtering criteria, and detailed coding guidelines for taxonomy development.

This substantial manual investment produced two key contributions: a comprehensive taxonomy grounded in systematic CVE mining and the first curated dataset of build-specific attacks, both essential resources for advancing build security research.

\section{Results}
We present our findings organized by research question. First, we introduce our taxonomy derived from CVE analysis. Second, we examine attack propagation patterns through representative cases. Finally, we validate the taxonomy against real-world incidents.

\subsection{Classification of Attack Vectors (RQ1)}
We derived a taxonomy of attack vectors based on the analysis of 621 CVEs, illustrated in Figure~\ref{fig:syntaxtree}. This taxonomy structures the classification according to three hierarchical dimensions:
\begin{enumerate}
    \item \textbf{Temporal (When)} -- pipeline stages where attacks occur
    \item \textbf{Technical (What)} -- exploitation mechanisms
    \item \textbf{Operational (How)} -- specific attack vectors
\end{enumerate}
 Each identified vector is associated with its potential impacts, forming a complete taxonomy.

The analysis reveals 15 distinct attack vectors distributed across 9 exploitation mechanisms over 5 CI/CD pipeline stages. The distribution shows an asymmetric concentration: Environment Setup presents 5 vectors (33\% of total), Dependency Resolution 4 vectors (27\%) and Compilation 3 vectors (20\%), together representing 80\% of identified vectors. This concentration on initial phases suggests that current security controls neglect these key entry points, requiring prioritization of protection measures.

A critical pattern emerges: specific mechanisms like Build Parameter Injection generate multiple distinct vectors (Remote Code Execution and Environment Variable Injection), while others in the taxonomy produce only one. This variability reveals fundamental differences in the exploitation potential of mechanisms.

\begin{flushleft}
\begin{tikzpicture}

\node [rqbox] (box){%
    \begin{minipage}{0.9\columnwidth}

\textbf{Classification of attack vectors.}  Our taxonomy identifies 14 distinct attack vectors organized through 8 exploitation mechanisms across 5 CI/CD pipeline stages. The classification reveals a critical concentration: 80\% of vectors (12 out of 15) target the first three phases (Environment Setup, Dependency Resolution, Compilation), with Environment Setup as the primary entry point (33\% of vectors). This When-What-How structure enables prioritization of security controls according to each pipeline phase's criticality.

    \end{minipage}
};
\node[titlerq, right=10pt] at (box.north west) {Answer to RQ1};
\end{tikzpicture}
\end{flushleft}

\subsection{Attack Propagation Across CI/CD Pipeline (RQ2)}
Analysis of our 40-attack dataset reveals two main categories of compromise: 23 destructive attacks (57.5\%) that contaminate build artifacts, and 17 extractive attacks (42.5\%) that exfiltrate sensitive information during the build process.

To understand the propagation mechanisms of these categories, we present a detailed analysis of two representative cases, selected for their comprehensive documentation and clear illustration of identified patterns:

\subsection*{Case 1: CrateDepression Attack -- Targeted Exploitation of GitLab CI}
The CrateDepression \cite{crateDepression-attack} incident exemplifies a destructive attack through sophisticated typosquatting attack in the Rust ecosystem. The attackers published a malicious package \texttt{rustdecimal}, mimicking the legitimate package \texttt{rust\_decimal}. The malicious code fully replicates the original library, except for the \texttt{Decimal::new} function which contains a conditional payload. This payload activates exclusively in GitLab CI environments through detection of the \texttt{GITLAB\_CI} environment variable.The attack was facilitated by impersonating a recognized Rust contributor, thus exploiting community trust to bypass detection mechanisms.

\textbf{Propagation Vector:}
\begin{enumerate}
    \item \textbf{Initial Phase:} Injection during dependency resolution
    \item \textbf{Execution Phase:} Activation of the malicious payload during the CI process
    \item \textbf{Distribution Phase:} Contamination of build artifacts destined for production
\end{enumerate}

This attack exploits two critical vulnerabilities: the implicit trust granted to open source packages and the difficulty of detecting subtle modifications in large codebases.

\subsection*{Case 2: Codecov Attack – CI/CD Chain Compromise via Secret Exfiltration}
The Codecov \cite{codecov-attack} incident exemplifies an extractive attack, highlights the impact of a vulnerability in a widely deployed third-party tool. Attackers exploited a misconfiguration in the Docker image of Codecov, exposing credentials that allowed them to modify the Bash Uploader script. This script, used to transmit code coverage reports to the Codecov platform, is typically executed during the testing phase in CI/CD pipelines.
\textbf{Attack Mechanism:}  
The attackers injected a malicious line of code into the Bash Uploader, enabling systematic exfiltration of environment variables to a server under their control. These variables often contained critical secrets: authentication tokens, API keys, Git credentials, and deployment identifiers.\\
\textbf{Propagation Vector:}
\begin{enumerate}
    \item \textbf{Testing Phase:} Execution of the modified Bash Uploader script
    \begin{itemize}
        \item Immediate exfiltration of environment variables
        \item Collection of secrets used throughout the pipeline
    \end{itemize}
    \item \textbf{Repository Compromise:} Exploitation of exfiltrated credentials
    \begin{itemize}
        \item Unauthorized access to private Git repositories
        \item Source code and commit history analysis
        \item Search for additional sensitive information
    \end{itemize}
    \item \textbf{Production Impact:} Escalation to critical environments
    \begin{itemize}
        \item Use of compromised production credentials
        \item Risk of lateral movement within the infrastructure
        \item Potential access to cloud and internal systems
    \end{itemize}
\end{enumerate}

This attack illustrates the critical importance of securing third-party tools integrated into CI/CD pipelines and the systemic risk posed by the compromise of a widely adopted component.\\

\begin{flushleft}
\begin{tikzpicture}

\node [rqbox] (box){%
    \begin{minipage}{0.9\columnwidth}
\textbf{Attack Propagation.}An analysis of real-world incidents affecting CI/CD systems reveals that attacks often propagate throughout the pipeline via compromised artifacts, shared environments, or reused credentials. This cascading propagation can lead to a full system compromise. We identify two main categories of attacks:
\begin{itemize}
    \item \textbf{Destructive attacks}, such as the \textit{CrateDepression Attack}, where build outputs are infected, compromising subsequent deployment stages.
    \item \textbf{Extractive attacks}, like the \textit{Codecov Attack}, which aim to exfiltrate sensitive information—such as credentials or API keys—by exploiting build scripts or environments.
\end{itemize}
These findings highlight the need for stronger security measures during the build phase, which is often overlooked despite being a critical component of the CI/CD pipeline.

    \end{minipage}
};
\node[titlerq, right=10pt] at (box.north west) {Answer to RQ2};
\end{tikzpicture}
\end{flushleft}

\subsection{Taxonomy Validation (RQ3)}
The proposed taxonomy is empirically validated through the analysis of 40 real attacks that occurred during the build phase, including 38 attacks extracted from the Atlantic Council dataset~\cite{atlantic_council_dataset} on software supply chain attacks and two identified through a literature review.

Table \ref{tab:taxonomy-validation} presents the validation of our taxonomy by mapping identified attack vectors to real-world attacks from our dataset. Of the 15 vectors in our taxonomy, eight are validated by at least one observed attack, demonstrating their empirical relevance. The seven unvalidated vectors (primarily in Environment Setup, Testing, and Packaging) represent potential threats identified through CVE analysis but not yet observed in documented attacks.

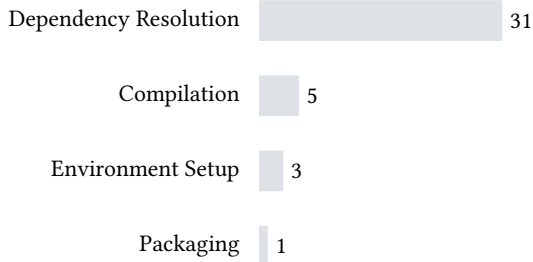
\begin{figure}[H]
\centering
\begin{tikzpicture}
\begin{axis}[ 
width=0.8\columnwidth,
xbar, xmin=0,xmax=50,  
%ytick pos=left,
%xtick pos=bottom,
%axis lines=none,
axis y line*=left,
axis line style={draw=none},
axis x line=none,
major y tick style={draw=none},
xlabel={Percentage \%},
symbolic y coords={Packaging, Environment Setup, Compilation, Dependency Resolution},
ytick=data,
nodes near coords, 
nodes near coords align={horizontal},
y=1cm,
]

\addplot [
    xbar,
    fill=crq2,
    draw=crq2,
    bar width=15pt
] coordinates {
    (1,{Packaging}) 
    (3,{Environment Setup}) 
    (5,{Compilation})
    (31,Dependency Resolution)};
\end{axis}
\end{tikzpicture}
\caption{\textbf{Distribution of Attacks Across Build Pipeline Stages} }
\label{fig:taxoval}
\end{figure}

Figure \ref{fig:taxoval} reveals an asymmetric distribution of attacks: Dependency Resolution concentrates 77.5\% of attacks (31/40), confirming this phase as attackers' preferred entry point. This dominance is explained by the diversity of exploitable vectors (Dependency Confusion, Typosquatting, Namespace Hijacking) and the difficulty of verifying the authenticity of all third-party dependencies.

\textit{Environment Setup (CI Pipeline Execution):} Although this phase contains five attack vectors in our taxonomy, only two are empirically validated (Secrets  Leakage and Malicious Payload Delivery) by three attacks, including the Codecov incident that compromised numerous organizations.

\textit{Dependency Resolution:} Three of the four identified vectors are validated by 31 attacks in this category. Dependency Confusion, Namespace Hijacking, and Typosquatting are all validated, with representative examples shown in Table \ref{tab:taxonomy-validation} including PyTorch, CrateDepression, JavaScript 2018 Backdoor, and RubyGems attacks among others.

\textit{Compilation:} With five documented attacks, this phase validates 2 of the three identified vectors: Code Injection (validated by SolarWinds Orion, XcodeGhost, Octopus Scanner, and XZ-utils) and Sensitive File Disclosure (validated by NotPetya). These attacks demonstrated the potential impact of compromise at this stage.

\clearpage

\begin{figure*}[htbp]
 \begin{adjustbox}{width=\textwidth, center}
   \begin{forest}
    for tree={
      draw, semithick, rounded corners, drop shadow,
       font=\normalsize,
       inner sep=7pt,        
    align=left,
    minimum width=30mm, 
           edge = {draw, semithick},
           anchor = east,
             grow = east,
    forked edge,            
            s sep = 4mm,    
            l sep = 8mm,    
         fork sep = 4mm,    % distance from parent to branching point
               }
[Build Process Compromise (CI Environment)\\, draw=blue, fill=blue!5 
    [Packaging\\, draw=blue, fill=blue!20
    [Trust Manipulation\\, draw=cyan, fill=cyan!20
    [Unsafe Deserialization\\ \small CVE-2021-21351 - CVE-2022-23630 - CVE-2023-41934\\, draw=teal, fill=teal!20
    [Unauthorized Modifications\\, draw=cyan, fill=cyan!5]]
    [Integrity Compromise\\ \small CVE-2019-10470\\, draw=teal, fill=teal!20
    [Unauthorized Modifications\\, draw=cyan, fill=cyan!5]] ]
    ]
    [Testing\\, draw=blue, fill=blue!20
     [Environment Manipulation\\, draw=cyan, fill=cyan!20
     [Variable Injection\\ \small CVE-2020-2256\\, draw=teal, fill=teal!20
        [Denial of Service\\, draw=cyan, fill=cyan!5] 
        [Arbitrary Code Execution\\, draw=cyan, fill=cyan!5] 
     ]
      ]
]
    [Compilation\\, draw=blue, fill=blue!20
         [Build Parameter Injection\\, draw=cyan, fill=cyan!20
         [Build Variable Leackage\\ \small CVE-2019-10358\\, draw=teal, fill=teal!20
         [Information Disclosure\\, draw=cyan, fill=cyan!5]]
         ]
         [Build Behavior Manipulation\\, draw=cyan, fill=cyan!20
          [Code Injection\\
         \small CVE-2020-11986 - \small CVE-2020-2211\\, draw=teal, fill=teal!20
         [Remote Command Execution\\, draw=cyan, fill=cyan!5]] 
         [Sensitive File Disclosure\\
         \small CVE-2023-35947 - CVE-2018-11804\\,draw=teal, fill=teal!20
         [Data Exfiltration\\, draw=cyan, fill=cyan!5]]
        ]
        ]
    [Dependency Resolution\\, draw=blue, fill=blue!20
        [Dependency Injection\\, draw=cyan, fill=cyan!20
         [MITM Attacks \\ \small CVE-2019-10240 - \small CVE-2021-41034\\,draw=teal, fill=teal!20
         [Malicious Code Injection\\, draw=cyan, fill=cyan!5]]
         [Namespace Hijacking \\ \small CVE-2021-26291 - \small CVE-2025-54313\\,draw=teal, fill=teal!20
         [Package Substitution\\, draw=cyan, fill=cyan!5]]
    [Typosquatting  \\ \small CVE-2017-16074 \\,draw=teal, fill=teal!20
    [Artifact Backdoor Injection\\, draw=cyan, fill=cyan!5]]
	[Dependency Confusion \\ \small CVE-2021-29427 - \small CVE-2021-29429 - \small CVE-2021-24105\\,draw=teal, fill=teal!20
    [Artifact Backdoor Injection\\, draw=cyan, fill=cyan!5]]
    ]
    ]
       [Environment Setup {(CI Pipeline Execution)}\\, draw=blue, fill=blue!20
        [Build Parameter Injection\\, draw=cyan, fill=cyan!20
        [Remote Code Execution \\ \small CVE-2024-27198 -  \small CVE-2023-52291 -  \small CVE-2021-15589\\,draw=teal, fill=teal!20
        [System Takeover\\, draw=cyan, fill=cyan!5]]
        [Environment Variable Injection \\ \small CVE-2021-32751\\,draw=teal, fill=teal!20
        [Build Manipulation\\, draw=cyan, fill=cyan!5]]
        ]
     [Secrets Exfiltration\\, draw=cyan, fill=cyan!20
        [Extraction of Authentication Secrets\\ \small CVE-2023-30853 - \small CVE-2019-10469 - \small CVE-2021-21361 - \small CVE-2021-32638\\
        - \small CVE-2023-30853 - \small CVE-2019-10413\\, draw=teal, fill=teal!20
     [Unauthorized Access\\, draw=cyan, fill=cyan!5]
     [Privilege Escalation \\, draw=cyan, fill=cyan!5]]
        ]
     [Lack of Access Control\\, draw=cyan, fill=cyan!20
     [Unauthorized Build Triggering\\ \small CVE-2020-2284 - CVE-2021-22186 - CVE-2018-20500 - CVE-2019-10324 - CVE-2020-2294\\, draw=teal, fill=teal!20
      [Pipeline Hijacking\\, draw=cyan, fill=cyan!5]]]
         [Improper Input Handling\\, draw=cyan, fill=cyan!20
     [Malicious Payload Delivery\\ \small CVE-2019-10327\\, draw=teal, fill=teal!20
     [Remote Code Execution\\, draw=cyan, fill=cyan!5]
     [Credential Theft\\, draw=cyan, fill=cyan!5]]]
    ]
]
\end{forest}
\end{adjustbox}
  \caption{Taxonomy of Attack Vectors Targeting the Build Process in CI/CD Environments}
  \caption*{Classification Based on Execution Stages in the Pipeline}
  \label{fig:syntaxtree}
    \vspace{1em}
\begin{tcolorbox}[colback=gray!5!white,colframe=black!50!gray!10,title=\textcolor{black!90}{Legend}]
\begin{tabular}{ll}
\fcolorbox{blue}{blue!20}{\textcolor{blue!20}{\rule{20pt}{7pt}}} & The build steps where the attack was initiated (Temporal) \\ \\
\fcolorbox{cyan}{cyan!20}{\textcolor{cyan!20}{\rule{20pt}{7pt}}} & Mechanism by which the attack occurred (Technical) \\ \\
\fcolorbox{teal}{teal!20}{\textcolor{teal!20}{\rule{20pt}{7pt}}} & The Attack Vector (Operational) \\ \\
\fcolorbox{cyan}{cyan!5}{\textcolor{cyan!5}{\rule{20pt}{7pt}}} & Example of attack impact \\ \\
\end{tabular}
\end{tcolorbox}
  \label{fig:syntaxtree}
\end{figure*}

\clearpage
\restoregeometry

\twocolumn

\textit{Packaging:} A single attack validates the Integrity Compromise vector, suggesting this final phase is either better protected or less attractive to attackers who favor upstream phases.

\textit{Testing:} No attacks in our dataset validate the Environment Manipulation vector identified in this phase, even though it is present in our derived taxonomy.

\begin{flushleft}
\begin{tikzpicture}

\node [rqbox] (box){%
    \begin{minipage}{0.9\columnwidth}
\textbf{Empirical Validation of the Taxonomy.} Analysis of 40 real attacks validates 8 of 15 identified vectors (53.3\% coverage). The asymmetric distribution (Table \ref{tab:taxonomy-validation} and Figure \ref{fig:taxoval}) reveals that Dependency Resolution constitutes the primary current attack vector (77.5\%), while Testing appears in no documented attacks. The seven unvalidated vectors, identified through CVE analysis, represent emerging or theoretical threats, highlighting our taxonomy's forward-looking value for anticipating attack evolution.

    \end{minipage}
};
\node[titlerq, right=10pt] at (box.north west) {Answer to RQ3};
\end{tikzpicture}
\end{flushleft}

\begin{table*}[ht!]
\centering
\caption{Validation of Taxonomy with Real-World Attack Vectors}
\begin{adjustbox}{width=\linewidth,center}

\label{tab:taxonomy-validation}

\begin{tabular}{lll}
\toprule
\textbf{Build Stage} & \textbf{Attack Vectors} & \textbf{Real-World Attacks } \\
\midrule

\textbf{Environment Setup (CI Pipeline Execution)} & 
\begin{tabular}[t]{@{}l@{}}
$\bullet$ Extraction of Authentication Secrets\\
$\bullet$ Malicious Payload Delivery\\
\end{tabular} & 
\begin{tabular}[t]{@{}l@{}}
$\bullet$ Heroku + Travis CI OAuth Token Theft \cite{heroku-attack}\\
$\bullet$ Codecov \cite{codecov-attack}, Nx Build \cite{nx-build-attack}\\
\end{tabular} \\
\midrule

\textbf{Dependency Resolution} & 
\begin{tabular}[t]{@{}l@{}}
$\bullet$ Dependency Confusion\\
$\bullet$ Namespace Hijacking\\
$\bullet$ Typosquatting\\
\quad\\
\end{tabular} & 
\begin{tabular}[t]{@{}l@{}}
$\bullet$ CrateDepression \cite{crateDepression-attack} , PyTorch \cite{pytorch-attack}\\
$\bullet$ JavaScript 2018 Backdoor \cite{java-script-backdoor-attack}, Webmin 1.890 Exploit \cite{webmin-attack}\\
$\bullet$ Azure npm \cite{azure-npm-attack}, bb-builder \cite{bb-builder-attack}\\
\quad RubyGems Jim Carrey and Peter Gibbons \cite{rubygems-attack}\\
\end{tabular} \\
\midrule

\textbf{Compilation} & 
\begin{tabular}[t]{@{}l@{}}
$\bullet$ Code Injection\\
\quad\\
$\bullet$ Sensitive File Disclosure\\

\end{tabular} & 
\begin{tabular}[t]{@{}l@{}}
$\bullet$ Octopus Scanner Malware \cite{octopus-attack}, XcodeGhost \cite{xcode-ghost-attack},\\
\quad SolarWinds Orion \cite{solarwinds-attack}, XZ-utils \cite{xz-utils-attack}\\
$\bullet$ NotPetya \cite{notpetya-attack}\\
\end{tabular} \\
\midrule

\textbf{Packaging} & 
\begin{tabular}[t]{@{}l@{}}
$\bullet$ Integrity Compromise\\
\end{tabular} & 
\begin{tabular}[t]{@{}l@{}}
$\bullet$ Java Code Signing Vulnerability \cite{java-code-signing-attack} \\
\end{tabular} \\
\bottomrule
\end{tabular}
\end{adjustbox}
\footnotesize{\textit{Note:} Representative examples from 40 real-world build attacks (2018-2024). For Dependency Resolution, 31 attacks are documented in our complete dataset.}

\end{table*}

%\twocolumn

\section{Discussion}
Our contribution addresses a gap identified in the literature: the absence of a systematic framework for characterizing threats specific to the build process, where build infrastructures represent the second major attack vector in software supply chain security \cite{software2025}.

Our results show that attacks are not uniformly distributed across the CI/CD pipeline. The Dependency Resolution phase accounts for 77.5\% of attacks in our dataset, confirming that reliance on public registries constitutes the most vulnerable point in the software supply chain. In contrast, Compilation attacks, although representing only 12.5\% of incidents, have a disproportionate impact as they compromise the final build output, affecting the entire deployment and distribution chain as well as end users. This demonstrates that criticality depends not only on frequency but also on scope and impact.

Environment Setup attacks reveal that secret management and malicious payload injection remain fundamental issues, confirming that security must be integrated from the initial configuration. Packaging emerges as an attack vector linked to artifact signing, a risk often underestimated in existing frameworks \cite{NIST-SP-800-204D}. The complete absence of Testing attacks despite identified vectors raises questions about whether this phase possesses inherent security properties or represents a research blind spot.

Our analysis further reveals that attacks rarely remain confined to a single stage; they propagate across the CI/CD pipeline through compromised artifacts, shared environments, or reused credentials. This cascading effect amplifies the initial compromise, potentially leading to full system subversion. The two dominant patterns—destructive attacks that infect build outputs and extractive attacks that exfiltrate sensitive information—underscore that securing individual steps in isolation is insufficient. Security controls must account for interdependencies between stages.

Unlike generic approaches, our taxonomy introduces vectors specific to Compilation and Packaging, which are often overlooked in current references. Each category is validated by real-world attacks, reinforcing empirical relevance. The taxonomy highlights the diversity and complexity of attack vectors, demonstrating that the build process security must be addressed holistically. Limiting efforts to dependency protection alone is insufficient.

This taxonomy provides a foundation for developing detection and classification models for build-specific CI/CD incidents and creating benchmarks to test security tools against build-related attack vectors. For practitioners, it offers guidance to prioritize security measures according to the most exposed build stages and design appropriate hardening policies for build environments.

\section{Background}
\begin{figure}[ht]
 \begin{adjustbox}{width=\linewidth, center}
 \includegraphics{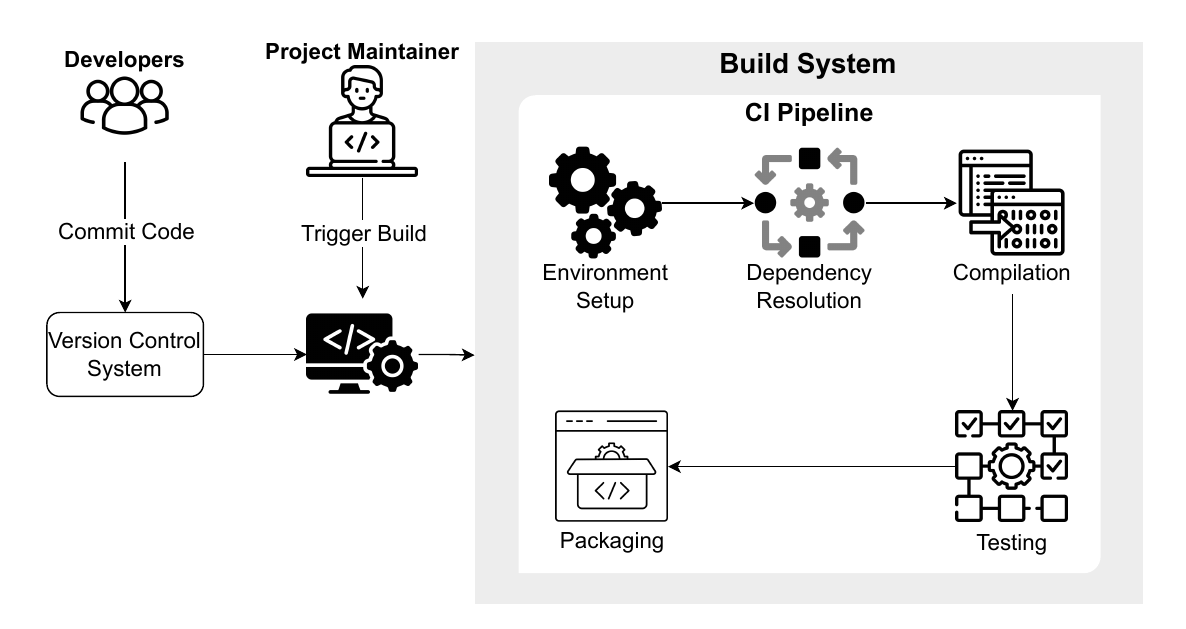}
\end{adjustbox}
\caption{\textbf{Build steps in a CI/CD environment}}
\label{fig:build_ste}
\end{figure}

\textit{\textbf{The Build Process as a Critical Attack Surface.}} Modern software development relies on complex build pipelines that transform source code and dependencies into deployable artifacts through automated stages including environment configuration, dependency resolution, compilation, and packaging.\\
Figure \ref{fig:build_ste} presents the standard stages of a build pipeline in a modern CI/CD environment, from developer to artifact registry. This model serves as the reference framework for our taxonomy, where each build stage (dependency resolution, compilation, packaging, deployment) represents a distinct category of attack vectors. While this automation enables efficient software delivery at scale, it creates an expanded attack surface vulnerable to sophisticated supply chain attacks \cite{236322}. The SolarWinds incident exemplified this vulnerability, demonstrating how build server compromise can inject malicious code into legitimately signed artifacts, bypassing traditional security controls even when source code remains pristine.\\

\textit{\textbf{Current Security Frameworks and Adoption Challenges.}} Industry frameworks such as SLSA (Supply chain Levels for Software Artifacts) and SSDF (Secure Software Development Framework) have emerged to address build security through practices including cryptographic signing, dependency validation, and reproducible builds. However, empirical evidence reveals limited adoption in open-source ecosystems. A recent industry study \cite{kalu2025industry} identified key barriers: implementation complexity, insufficient automation tooling, organizational costs, and low security awareness among maintainers. This adoption gap is particularly concerning given the escalating threat landscape.\\

\textit{\textbf{Empirical Evidence of Growing Threats.}} Supply chain attacks targeting build processes have increased dramatically, with measurable strategic impact across the software ecosystem. The Atlantic Council's comprehensive dataset documents over 168 incidents from the past decade, providing rich empirical data on attack vectors, timelines, and impacts 
\cite{atlantic_council_dataset}. Yet despite this wealth of incident data, the research community lacks empirically-validated taxonomies specifically addressing build-phase vulnerabilities—a critical gap that impedes systematic defense design.\\

\section{Related Work}
\paragraph{\textbf{Mining Software Repositories for Security.}}
Mining software repositories is a widely used approach for studying security 
vulnerabilities. Wang et al. \cite{10.1145/3639478.3647634} analyzed 6,134 
CVEs to construct ReposVul, a dataset for training detection models. While 
their methodology demonstrates the value of large-scale CVE analysis, they 
focus on general vulnerabilities without distinguishing build-specific threats. 
Akhoundali et al. \cite{10.1145/3663533.3664036} proposed MoreFixes, built 
from 77,597 CVEs mapped to GitHub projects, creating a comprehensive 
vulnerability-fix dataset. These studies validate CVE analysis for improving 
software security but treat build vulnerabilities as part of general security 
issues rather than a distinct category requiring specialized analysis.

\paragraph{\textbf{CI/CD Pipeline Security.}}
Recent research has begun examining build infrastructure vulnerabilities 
specifically. \cite{281396} characterized the security of GitHub 
CI workflows, revealing critical vulnerabilities in workflow configurations 
and script execution. Building on this, \cite{291265} developed Argus, 
a static taint analysis system that analyzed 2.8 million workflows across 
1 million repositories, detecting high and medium impact code injection 
vulnerabilities in 4,307 workflows. \cite{10179471} conducted a systematic 
study across seven major CI platforms, uncovering novel threats including 
task hijacking, repository privilege escalation, and artifact hijacking. 
These works demonstrate the unique security challenges in build infrastructures 
but focus on specific platforms or vulnerability types rather than providing 
a comprehensive taxonomy across all build phases.

\paragraph{\textbf{Software Supply Chain Security.}}
Parallel research explores attacks targeting the software supply chain. 
Williams et al. \cite{software2025} address security issues in the software 
supply chain, identifying build infrastructures as the second major attack 
vector, yet their treatment of build-specific mechanisms remains high-level. 
Neupane et al. \cite{neupane2023beyond} analyze package confusion attacks, 
an emerging form of software supply chain attack that exploits dependency 
resolution—a critical build phase. Yet their analysis focuses on this single 
vector without considering the broader build attack surface. Ladisa et al. 
\cite{ladisa2023sok} proposed a taxonomy of attacks targeting the open source 
software supply chain based on analysis of real attacks and scientific 
literature. While comprehensive with their attack vectors linked to real-world 
incidents, their taxonomy does not detail build-specific mechanisms or 
distinguish between different build phases. \cite{ohm2020backstabber} reviewed open 
source supply chain attacks in the "Backstabber's Knife Collection," providing 
valuable insights into attack patterns but without the build-phase granularity 
needed for targeted defenses.

\paragraph{\textbf{Research Gap.}}
These works have several limitations. First, existing mining studies have not 
specifically analyzed build-related CVEs to understand their unique 
characteristics \cite{10.1145/3639478.3647634, 10.1145/3663533.3664036}. 
Second, while CI/CD security research \cite{281396, 291265, 
10179471} provides valuable insights into specific platforms and vulnerability 
types, no work systematically maps attack vectors across all build phases. 
Third, supply chain security frameworks treat the build process superficially, 
lacking the granularity to distinguish vulnerabilities across build phases 
\cite{software2025, neupane2023beyond, ladisa2023sok}. No existing taxonomy 
provides systematic classification of attack vectors specific to different 
build stages—environment setup, dependency resolution, compilation, testing, 
and packaging. Our work addresses these gaps by analyzing 621 build-specific 
CVEs, constructing a hierarchical taxonomy mapping attack vectors to specific 
build phases, and validating against 40 real-world build attacks. This provides 
the first comprehensive framework dedicated to build process security.

\section{Threat to Validity}
\paragraph{Taxonomy Construction Process.} 
Our categorization of attack vectors relies on manual interpretation of injection points within the build pipeline. To mitigate subjectivity, we employed a systematic classification protocol with inclusion/exclusion criteria. However, some vulnerabilities could reasonably belong to multiple categories, and alternative taxonomies could emerge from different conceptual frameworks.

\paragraph{CVE Mining Limitations.} 
Although we extracted 621 CVEs using rigorous selection criteria, the level of detail in CVE descriptions and documentation varies. Build-related vulnerabilities may be underrepresented due to incomplete metadata and ambiguous descriptions. Furthermore, our keyword-based filtering approach may have missed relevant CVEs using non-standard terminology.

\paragraph{Validation Dataset Scope.} 
Our validation relies on 38 build process attacks derived from analyzing 168 supply chain attacks from the Atlantic Council dataset, plus two recent incidents from literature review. While this represents the most comprehensive publicly available compilation, it inherently excludes: attacks not disclosed due to legal or reputational concerns, and sophisticated attacks that remain undetected. Moreover, the level of detail in attack descriptions varies, with some preventing precise mapping to build processes.

\paragraph{Temporal Validity.} 
The rapidly evolving build ecosystem---including emerging technologies such as AI-assisted build tools, containerized build environments, and novel dependency management systems---could introduce attack vectors not yet represented in our framework. Our snapshot analysis may not capture emerging trends or obsolete attack patterns. Regular taxonomy updates will be necessary to maintain relevance, particularly as new build technologies and security practices emerge.

\section{Data Availability} 
Our dataset is publicly available at \url{https://anonymous.4open.science/r/Taxonomizing-Build-Attacks-8BB0}, making our results reproducible.
\section{Conclusion and Future Work}
\subsection{Conclusion} 
Our empirical analysis of 621 build-specific CVEs and 40 documented build process attacks establishes a comprehensive taxonomy of attack vectors across the software build pipeline. The proposed taxonomy reveals 15 distinct attack vectors distributed across five build phases—Environment Setup, Dependency Resolution, Compilation, Testing, and Packaging—extending far beyond traditional security approaches focused solely on protecting compilation tools. 

Our characterization identifies two dominant propagation patterns with distinct security implications: destructive attacks that contaminate build artifacts for downstream propagation, and extractive attacks that exploit build environments for credential exfiltration. The concentration of 77.5\% of attacks on Dependency Resolution, combined with the disproportionate impact of Compilation attacks (12.5\% frequency but system-wide consequences), demonstrates that both frequency and impact must guide security prioritization. 

Empirical validation reveals a significant gap: only 8 of 15 identified vectors (53.3\%) appear in real-world attacks. Seven vectors—Remote Code Execution, Environment Variable Injection, Unauthorized Build Triggering, MITM Attacks, Build Variable Leakage, Variable Injection, and Unsafe Deserialization—remain unvalidated, representing either unexploited vulnerabilities or emerging threats. This divergence between theoretical vulnerabilities and practical exploitation underscores the evolving nature of build security threats.

\subsection{Future Work}
These findings highlight critical research directions. First, investigating why certain vectors remain unexploited could reveal defensive opportunities or predict future attack evolution. Second, developing automated detection tools specifically calibrated to the validated attack vectors would provide immediate security value. Third, understanding the complete absence of Testing phase attacks despite identified vulnerabilities warrants deeper investigation. 

For practitioners, our taxonomy enables systematic integration of build security into DevSecOps practices through phase-specific controls and targeted hardening policies. The identification of cross-cutting mechanisms like Build Parameter Injection appearing in multiple phases suggests architectural vulnerabilities requiring fundamental redesign rather than point solutions. 

As software supply chain attacks grow increasingly sophisticated, securing the build process becomes essential. Our taxonomy provides the foundation for transforming build security from an afterthought to a systematic discipline.

\clearpage

% \bibliographystyle{ACM-Reference-Format}
% \bibliography{main}

\begin{thebibliography}{35}

%%% the Web address.

\ifx \showCODEN    \undefined \def \showCODEN     #1{\unskip}     \fi
\ifx \showISBNx    \undefined \def \showISBNx     #1{\unskip}     \fi
\ifx \showISBNxiii \undefined \def \showISBNxiii  #1{\unskip}     \fi
\ifx \showISSN     \undefined \def \showISSN      #1{\unskip}     \fi
\ifx \showLCCN     \undefined \def \showLCCN      #1{\unskip}     \fi
\ifx \shownote     \undefined \def \shownote      #1{#1}          \fi
\ifx \showarticletitle \undefined \def \showarticletitle #1{#1}   \fi
\ifx \showURL      \undefined \def \showURL       {\relax}        \fi
% invisible to TeX
\providecommand\bibfield[2]{#2}
\providecommand\bibinfo[2]{#2}
\providecommand\natexlab[1]{#1}
\providecommand\showeprint[2][]{arXiv:#2}

\bibitem[pyt(2013)]%
        {pytorch-attack}
 \bibinfo{year}{2013}\natexlab{}.
\newblock \bibinfo{booktitle}{\emph{PyTorch discloses malicious dependency chain compromise over holidays}}.
\newblock
\urldef\tempurl%
\url{https://www.bleepingcomputer.com/news/security/pytorch-discloses-malicious-dependency-chain-compromise-over-holidays/}
\showURL{%
Retrieved September 11, 2025 from \tempurl}


\bibitem[xco(2015)]%
        {xcode-ghost-attack}
 \bibinfo{year}{2015}\natexlab{}.
\newblock \bibinfo{booktitle}{\emph{Apple's App Store infected with XcodeGhost malware}}.
\newblock
\urldef\tempurl%
\url{https://www.bbc.com/news/technology-34311203}
\showURL{%
Retrieved September 04, 2025 from \tempurl}


\bibitem[cra(2022)]%
        {crateDepression-attack}
 \bibinfo{year}{2022}\natexlab{}.
\newblock \bibinfo{booktitle}{\emph{malicious package rustdecimal}}.
\newblock
\urldef\tempurl%
\url{https://github.com/paupino/rust-decimal/issues/514#issuecomment-1115456464,%20https://www.sentinelone.com/labs/cratedepression-rust-supply-chain-attack-infects-cloud-ci-pipelines-with-go-malware/,%20https://thehackernews.com/2022/05/researchers-uncover-rust-supply-chain.html}
\showURL{%
Retrieved September 11, 2025 from \tempurl}


\bibitem[xz-(2024)]%
        {xz-utils-attack}
 \bibinfo{year}{2024}\natexlab{}.
\newblock \bibinfo{booktitle}{\emph{XZ Utils Backdoor — Everything You Need to Know, and What You Can Do}}.
\newblock
\urldef\tempurl%
\url{https://www.akamai.com/blog/security-research/critical-linux-backdoor-xz-utils-discovered-what-to-know}
\showURL{%
Retrieved September 04, 2025 from \tempurl}


\bibitem[Akhoundali et~al\mbox{.}(2024)]%
        {10.1145/3663533.3664036}
\bibfield{author}{\bibinfo{person}{Jafar Akhoundali}, \bibinfo{person}{Sajad~Rahim Nouri}, \bibinfo{person}{Kristian Rietveld}, {and} \bibinfo{person}{Olga Gadyatskaya}.} \bibinfo{year}{2024}\natexlab{}.
\newblock \showarticletitle{MoreFixes: A Large-Scale Dataset of CVE Fix Commits Mined through Enhanced Repository Discovery}. In \bibinfo{booktitle}{\emph{Proceedings of the 20th International Conference on Predictive Models and Data Analytics in Software Engineering}} (Porto de Galinhas, Brazil) \emph{(\bibinfo{series}{PROMISE 2024})}. \bibinfo{publisher}{Association for Computing Machinery}, \bibinfo{address}{New York, NY, USA}, \bibinfo{pages}{42–51}.
\newblock
\showISBNx{9798400706752}
\href{https://doi.org/10.1145/3663533.3664036}{doi:\nolinkurl{10.1145/3663533.3664036}}


\bibitem[Cameron(2024)]%
        {webmin-attack}
\bibfield{author}{\bibinfo{person}{Jamie Cameron}.} \bibinfo{year}{2024}\natexlab{}.
\newblock \bibinfo{booktitle}{\emph{Webmin}}.
\newblock
\urldef\tempurl%
\url{http://www.webmin.com/exploit.html}
\showURL{%
Retrieved September 04, 2025 from \tempurl}


\bibitem[Catalin~Cimpanu(2018)]%
        {java-script-backdoor-attack}
\bibfield{author}{\bibinfo{person}{Contributor Catalin~Cimpanu}.} \bibinfo{year}{2018}\natexlab{}.
\newblock \bibinfo{booktitle}{\emph{Hacker backdoors popular JavaScript library to steal Bitcoin funds}}.
\newblock
\urldef\tempurl%
\url{https://www.zdnet.com/article/hacker-backdoors-popular-javascript-library-to-steal-bitcoin-funds/}
\showURL{%
Retrieved September 11, 2025 from \tempurl}


\bibitem[Chandramouli et~al\mbox{.}(2024)]%
        {NIST-SP-800-204D}
\bibfield{author}{\bibinfo{person}{Ramaswamy Chandramouli}, \bibinfo{person}{Frederick Kautz}, {and} \bibinfo{person}{Santiago Torres-Arias}.} \bibinfo{year}{2024}\natexlab{}.
\newblock \bibinfo{booktitle}{\emph{Strategies for the Integration of Software Supply Chain Security in DevSecOps CI/CD Pipelines}}.
\newblock \bibinfo{type}{{T}echnical {R}eport} NIST Special Publication (SP) 800-204D. \bibinfo{institution}{National Institute of Standards and Technology}, \bibinfo{address}{Gaithersburg, MD}.
\newblock
\href{https://doi.org/10.6028/NIST.SP.800-204D}{doi:\nolinkurl{10.6028/NIST.SP.800-204D}}


\bibitem[Constantin(2020)]%
        {solarwinds-attack}
\bibfield{author}{\bibinfo{person}{Lucian Constantin}.} \bibinfo{year}{2020}\natexlab{}.
\newblock \bibinfo{booktitle}{\emph{SolarWinds attack explained: And why it was so hard to detect}}.
\newblock
\urldef\tempurl%
\url{https://www.csoonline.com/article/570191/solarwinds-supply-chain-attack-explained-why-organizations-were-not-prepared.html}
\showURL{%
Retrieved September 04, 2025 from \tempurl}


\bibitem[Council(2020)]%
        {atlantic_council_dataset}
\bibfield{author}{\bibinfo{person}{Atlantic Council}.} \bibinfo{year}{2020}\natexlab{}.
\newblock \bibinfo{booktitle}{\emph{Software supply chain security: The dataset}}.
\newblock
\urldef\tempurl%
\url{https://www.atlanticcouncil.org/content-series/cybersecurity-policy-and-strategy/software-supply-chain-security-the-dataset/}
\showURL{%
Retrieved October 15, 2025 from \tempurl}


\bibitem[Debian(2019)]%
        {dl1}
\bibfield{author}{\bibinfo{person}{Debian}.} \bibinfo{year}{2019}\natexlab{}.
\newblock \bibinfo{booktitle}{\emph{Reproducible Builds}}.
\newblock
\urldef\tempurl%
\url{https://wiki.debian.org/ReproducibleBuilds}
\showURL{%
Retrieved October 21, 2025 from \tempurl}


\bibitem[Debian(2024)]%
        {dl2}
\bibfield{author}{\bibinfo{person}{Debian}.} \bibinfo{year}{2024}\natexlab{}.
\newblock \bibinfo{booktitle}{\emph{Reproducible Builds}}.
\newblock
\urldef\tempurl%
\url{https://reproducible-builds.org/}
\showURL{%
Retrieved October 21, 2025 from \tempurl}


\bibitem[FireEye(2020)]%
        {SUNBURST_backdoor}
\bibfield{author}{\bibinfo{person}{FireEye}.} \bibinfo{year}{2020}\natexlab{}.
\newblock \bibinfo{booktitle}{\emph{Highly Evasive Attacker Leverages SolarWinds Supply Chain to Compromise Multiple Global Victims With SUNBURST Backdoor}}.
\newblock
\urldef\tempurl%
\url{https://cloud.google.com/blog/topics/threat-intelligence/evasive-attacker-leverages-solarwinds-supply-chain-compromises-with-sunburst-backdoor/}
\showURL{%
Retrieved September 20, 2025 from \tempurl}


\bibitem[Github(2020)]%
        {bb-builder-attack}
\bibfield{author}{\bibinfo{person}{Github}.} \bibinfo{year}{2020}\natexlab{}.
\newblock \bibinfo{booktitle}{\emph{Malicious Package in bb-builder}}.
\newblock
\urldef\tempurl%
\url{https://github.com/advisories/GHSA-vm6v-w6q2-mrrq}
\showURL{%
Retrieved September 04, 2025 from \tempurl}


\bibitem[GOODIN(2020)]%
        {rubygems-attack}
\bibfield{author}{\bibinfo{person}{Dan GOODIN}.} \bibinfo{year}{2020}\natexlab{}.
\newblock \bibinfo{booktitle}{\emph{Supply-chain attack hits RubyGems repository with 725 malicious packages}}.
\newblock
\urldef\tempurl%
\url{https://arstechnica.com/information-technology/2020/04/725-bitcoin-stealing-apps-snuck-into-ruby-repository/}
\showURL{%
Retrieved September 04, 2025 from \tempurl}


\bibitem[GOODIN(2024)]%
        {notpetya-attack}
\bibfield{author}{\bibinfo{person}{Dan GOODIN}.} \bibinfo{year}{2024}\natexlab{}.
\newblock \bibinfo{booktitle}{\emph{Backdoor built in to widely used tax app seeded last week’s NotPetya outbreak}}.
\newblock
\urldef\tempurl%
\url{https://arstechnica.com/information-technology/2017/07/heavily-armed-police-raid-company-that-seeded-last-weeks-notpetya-outbreak/}
\showURL{%
Retrieved September 04, 2025 from \tempurl}


\bibitem[Gu et~al\mbox{.}(2023)]%
        {10179471}
\bibfield{author}{\bibinfo{person}{Yacong Gu}, \bibinfo{person}{Lingyun Ying}, \bibinfo{person}{Huajun Chai}, \bibinfo{person}{Chu Qiao}, \bibinfo{person}{Haixin Duan}, {and} \bibinfo{person}{Xing Gao}.} \bibinfo{year}{2023}\natexlab{}.
\newblock \showarticletitle{Continuous Intrusion: Characterizing the Security of Continuous Integration Services}. In \bibinfo{booktitle}{\emph{2023 IEEE Symposium on Security and Privacy (SP)}}. \bibinfo{pages}{1561--1577}.
\newblock
\href{https://doi.org/10.1109/SP46215.2023.10179471}{doi:\nolinkurl{10.1109/SP46215.2023.10179471}}


\bibitem[Kalu et~al\mbox{.}(2025)]%
        {kalu2025industry}
\bibfield{author}{\bibinfo{person}{Kelechi~G Kalu}, \bibinfo{person}{Tanmay Singla}, \bibinfo{person}{Chinenye Okafor}, \bibinfo{person}{Santiago Torres-Arias}, {and} \bibinfo{person}{James~C Davis}.} \bibinfo{year}{2025}\natexlab{}.
\newblock \showarticletitle{An industry interview study of software signing for supply chain security}. In \bibinfo{booktitle}{\emph{34th USENIX Security Symposium (USENIX Security 25)}}. \bibinfo{pages}{81--100}.
\newblock


\bibitem[Koishybayev et~al\mbox{.}(2022)]%
        {281396}
\bibfield{author}{\bibinfo{person}{Igibek Koishybayev}, \bibinfo{person}{Aleksandr Nahapetyan}, \bibinfo{person}{Raima Zachariah}, \bibinfo{person}{Siddharth Muralee}, \bibinfo{person}{Bradley Reaves}, \bibinfo{person}{Alexandros Kapravelos}, {and} \bibinfo{person}{Aravind Machiry}.} \bibinfo{year}{2022}\natexlab{}.
\newblock \showarticletitle{Characterizing the Security of Github {CI} Workflows}. In \bibinfo{booktitle}{\emph{31st USENIX Security Symposium (USENIX Security 22)}}. \bibinfo{publisher}{USENIX Association}, \bibinfo{address}{Boston, MA}, \bibinfo{pages}{2747--2763}.
\newblock
\showISBNx{978-1-939133-31-1}
\urldef\tempurl%
\url{https://www.usenix.org/conference/usenixsecurity22/presentation/koishybayev}
\showURL{%
\tempurl}


\bibitem[Ladisa et~al\mbox{.}(2023)]%
        {ladisa2023sok}
\bibfield{author}{\bibinfo{person}{Piergiorgio Ladisa}, \bibinfo{person}{Henrik Plate}, \bibinfo{person}{Matias Martinez}, {and} \bibinfo{person}{Olivier Barais}.} \bibinfo{year}{2023}\natexlab{}.
\newblock \showarticletitle{SoK: Taxonomy of Attacks on Open-Source Software Supply Chains}. In \bibinfo{booktitle}{\emph{2023 IEEE Symposium on Security and Privacy (SP)}}. \bibinfo{pages}{1509--1526}.
\newblock
\href{https://doi.org/10.1109/SP46215.2023.10179304}{doi:\nolinkurl{10.1109/SP46215.2023.10179304}}


\bibitem[Lakshmanan(2024)]%
        {azure-npm-attack}
\bibfield{author}{\bibinfo{person}{Ravie Lakshmanan}.} \bibinfo{year}{2024}\natexlab{}.
\newblock \bibinfo{booktitle}{\emph{Over 200 Malicious NPM Packages Caught Targeting Azure Developers}}.
\newblock
\urldef\tempurl%
\url{https://thehackernews.com/2022/03/over-200-malicious-npm-packages-caught.html}
\showURL{%
Retrieved September 04, 2025 from \tempurl}


\bibitem[Leyden(2020)]%
        {octopus-attack}
\bibfield{author}{\bibinfo{person}{John Leyden}.} \bibinfo{year}{2020}\natexlab{}.
\newblock \bibinfo{booktitle}{\emph{How Octopus Scanner malware attacked the open source supply chain}}.
\newblock
\urldef\tempurl%
\url{https://portswigger.net/daily-swig/how-octopus-scanner-malware-attacked-the-open-source-supply-chain}
\showURL{%
Retrieved September 04, 2025 from \tempurl}


\bibitem[Muralee et~al\mbox{.}(2023)]%
        {291265}
\bibfield{author}{\bibinfo{person}{Siddharth Muralee}, \bibinfo{person}{Igibek Koishybayev}, \bibinfo{person}{Aleksandr Nahapetyan}, \bibinfo{person}{Greg Tystahl}, \bibinfo{person}{Brad Reaves}, \bibinfo{person}{Antonio Bianchi}, \bibinfo{person}{William Enck}, \bibinfo{person}{Alexandros Kapravelos}, {and} \bibinfo{person}{Aravind Machiry}.} \bibinfo{year}{2023}\natexlab{}.
\newblock \showarticletitle{{ARGUS}: A Framework for Staged Static Taint Analysis of {GitHub} Workflows and Actions}. In \bibinfo{booktitle}{\emph{32nd USENIX Security Symposium (USENIX Security 23)}}. \bibinfo{publisher}{USENIX Association}, \bibinfo{address}{Anaheim, CA}, \bibinfo{pages}{6983--7000}.
\newblock
\showISBNx{978-1-939133-37-3}
\urldef\tempurl%
\url{https://www.usenix.org/conference/usenixsecurity23/presentation/muralee}
\showURL{%
\tempurl}


\bibitem[Neupane et~al\mbox{.}(2023)]%
        {neupane2023beyond}
\bibfield{author}{\bibinfo{person}{Shradha Neupane}, \bibinfo{person}{Grant Holmes}, \bibinfo{person}{Elizabeth Wyss}, \bibinfo{person}{Drew Davidson}, {and} \bibinfo{person}{Lorenzo De~Carli}.} \bibinfo{year}{2023}\natexlab{}.
\newblock \showarticletitle{Beyond typosquatting: an in-depth look at package confusion}. In \bibinfo{booktitle}{\emph{32nd USENIX Security Symposium (USENIX Security 23)}}. \bibinfo{pages}{3439--3456}.
\newblock


\bibitem[Ohm et~al\mbox{.}(2020)]%
        {ohm2020backstabber}
\bibfield{author}{\bibinfo{person}{Marc Ohm}, \bibinfo{person}{Henrik Plate}, \bibinfo{person}{Arnold Sykosch}, {and} \bibinfo{person}{Michael Meier}.} \bibinfo{year}{2020}\natexlab{}.
\newblock \showarticletitle{Backstabber’s knife collection: A review of open source software supply chain attacks}. In \bibinfo{booktitle}{\emph{Detection of Intrusions and Malware, and Vulnerability Assessment: 17th International Conference, DIMVA 2020, Lisbon, Portugal, June 24--26, 2020, Proceedings 17}}. Springer, \bibinfo{pages}{23--43}.
\newblock


\bibitem[Osborne(2021)]%
        {codecov-attack}
\bibfield{author}{\bibinfo{person}{Charlie Osborne}.} \bibinfo{year}{2021}\natexlab{}.
\newblock \bibinfo{booktitle}{\emph{Codecov breach impacted ‘hundreds’ of customer networks: report}}.
\newblock
\urldef\tempurl%
\url{https://www.zdnet.com/article/codecov-breach-impacted-hundreds-of-customer-networks/}
\showURL{%
Retrieved September 11, 2025 from \tempurl}


\bibitem[Peisert et~al\mbox{.}(2021)]%
        {9382367}
\bibfield{author}{\bibinfo{person}{Sean Peisert}, \bibinfo{person}{Bruce Schneier}, \bibinfo{person}{Hamed Okhravi}, \bibinfo{person}{Fabio Massacci}, \bibinfo{person}{Terry Benzel}, \bibinfo{person}{Carl Landwehr}, \bibinfo{person}{Mohammad Mannan}, \bibinfo{person}{Jelena Mirkovic}, \bibinfo{person}{Atul Prakash}, {and} \bibinfo{person}{James~Bret Michael}.} \bibinfo{year}{2021}\natexlab{}.
\newblock \showarticletitle{Perspectives on the SolarWinds Incident}.
\newblock \bibinfo{journal}{\emph{IEEE Security \& Privacy}} \bibinfo{volume}{19}, \bibinfo{number}{2} (\bibinfo{year}{2021}), \bibinfo{pages}{7--13}.
\newblock
\href{https://doi.org/10.1109/MSEC.2021.3051235}{doi:\nolinkurl{10.1109/MSEC.2021.3051235}}


\bibitem[Ren et~al\mbox{.}(2018)]%
        {10.1145/3180155.3180224}
\bibfield{author}{\bibinfo{person}{Zhilei Ren}, \bibinfo{person}{He Jiang}, \bibinfo{person}{Jifeng Xuan}, {and} \bibinfo{person}{Zijiang Yang}.} \bibinfo{year}{2018}\natexlab{}.
\newblock \showarticletitle{Automated localization for unreproducible builds}. In \bibinfo{booktitle}{\emph{Proceedings of the 40th International Conference on Software Engineering}} (Gothenburg, Sweden) \emph{(\bibinfo{series}{ICSE '18})}. \bibinfo{publisher}{Association for Computing Machinery}, \bibinfo{address}{New York, NY, USA}, \bibinfo{pages}{71–81}.
\newblock
\showISBNx{9781450356381}
\href{https://doi.org/10.1145/3180155.3180224}{doi:\nolinkurl{10.1145/3180155.3180224}}


\bibitem[Ren et~al\mbox{.}(2022)]%
        {10.1145/3510003.3510102}
\bibfield{author}{\bibinfo{person}{Zhilei Ren}, \bibinfo{person}{Shiwei Sun}, \bibinfo{person}{Jifeng Xuan}, \bibinfo{person}{Xiaochen Li}, \bibinfo{person}{Zhide Zhou}, {and} \bibinfo{person}{He Jiang}.} \bibinfo{year}{2022}\natexlab{}.
\newblock \showarticletitle{Automated patching for unreproducible builds}. In \bibinfo{booktitle}{\emph{Proceedings of the 44th International Conference on Software Engineering}} (Pittsburgh, Pennsylvania) \emph{(\bibinfo{series}{ICSE '22})}. \bibinfo{publisher}{Association for Computing Machinery}, \bibinfo{address}{New York, NY, USA}, \bibinfo{pages}{200–211}.
\newblock
\showISBNx{9781450392211}
\href{https://doi.org/10.1145/3510003.3510102}{doi:\nolinkurl{10.1145/3510003.3510102}}


\bibitem[Swain(2025)]%
        {nx-build-attack}
\bibfield{author}{\bibinfo{person}{Gyana Swain}.} \bibinfo{year}{2025}\natexlab{}.
\newblock \bibinfo{booktitle}{\emph{Nx build Attack}}.
\newblock
\urldef\tempurl%
\url{https://www.lemondeinformatique.fr/actualites/lire-la-poste-applique-la-criminalistique-a-la-lutte-contre-la-fraude-98126.html/}
\showURL{%
Retrieved September 11, 2025 from \tempurl}


\bibitem[Torres-Arias et~al\mbox{.}(2019)]%
        {236322}
\bibfield{author}{\bibinfo{person}{Santiago Torres-Arias}, \bibinfo{person}{Hammad Afzali}, \bibinfo{person}{Trishank~Karthik Kuppusamy}, \bibinfo{person}{Reza Curtmola}, {and} \bibinfo{person}{Justin Cappos}.} \bibinfo{year}{2019}\natexlab{}.
\newblock \showarticletitle{in-toto: Providing farm-to-table guarantees for bits and bytes}. In \bibinfo{booktitle}{\emph{28th USENIX Security Symposium (USENIX Security 19)}}. \bibinfo{publisher}{USENIX Association}, \bibinfo{address}{Santa Clara, CA}, \bibinfo{pages}{1393--1410}.
\newblock
\showISBNx{978-1-939133-06-9}
\urldef\tempurl%
\url{https://www.usenix.org/conference/usenixsecurity19/presentation/torres-arias}
\showURL{%
\tempurl}


\bibitem[Tung(2013)]%
        {java-code-signing-attack}
\bibfield{author}{\bibinfo{person}{Liam Tung}.} \bibinfo{year}{2013}\natexlab{}.
\newblock \bibinfo{booktitle}{\emph{Java zero-day malware 'was signed with certificates stolen from security vendor'}}.
\newblock
\urldef\tempurl%
\url{https://www.zdnet.com/article/java-zero-day-malware-was-signed-with-certificates-stolen-from-security-vendor/}
\showURL{%
Retrieved September 06, 2025 from \tempurl}


\bibitem[Vijayan(2022)]%
        {heroku-attack}
\bibfield{author}{\bibinfo{person}{Jai Vijayan}.} \bibinfo{year}{2022}\natexlab{}.
\newblock \bibinfo{booktitle}{\emph{Heroku: Cyberattacker Used Stolen OAuth Tokens to Steal Customer Account Credentials}}.
\newblock
\urldef\tempurl%
\url{https://www.darkreading.com/endpoint/heroku-cyberattacker-stolen-oauth-token-customer-account-credentials}
\showURL{%
Retrieved September 11, 2025 from \tempurl}


\bibitem[Wang et~al\mbox{.}(2024)]%
        {10.1145/3639478.3647634}
\bibfield{author}{\bibinfo{person}{Xinchen Wang}, \bibinfo{person}{Ruida Hu}, \bibinfo{person}{Cuiyun Gao}, \bibinfo{person}{Xin-Cheng Wen}, \bibinfo{person}{Yujia Chen}, {and} \bibinfo{person}{Qing Liao}.} \bibinfo{year}{2024}\natexlab{}.
\newblock \showarticletitle{ReposVul: A Repository-Level High-Quality Vulnerability Dataset}. In \bibinfo{booktitle}{\emph{Proceedings of the 2024 IEEE/ACM 46th International Conference on Software Engineering: Companion Proceedings}} (Lisbon, Portugal) \emph{(\bibinfo{series}{ICSE-Companion '24})}. \bibinfo{publisher}{Association for Computing Machinery}, \bibinfo{address}{New York, NY, USA}, \bibinfo{pages}{472–483}.
\newblock
\showISBNx{9798400705021}
\href{https://doi.org/10.1145/3639478.3647634}{doi:\nolinkurl{10.1145/3639478.3647634}}


\bibitem[Williams et~al\mbox{.}(2025)]%
        {software2025}
\bibfield{author}{\bibinfo{person}{Laurie Williams}, \bibinfo{person}{Giacomo Benedetti}, \bibinfo{person}{Sivana Hamer}, \bibinfo{person}{Ranindya Paramitha}, \bibinfo{person}{Imranur Rahman}, \bibinfo{person}{Mahzabin Tamanna}, \bibinfo{person}{Greg Tystahl}, \bibinfo{person}{Nusrat Zahan}, \bibinfo{person}{Patrick Morrison}, \bibinfo{person}{Yasemin Acar}, \bibinfo{person}{Michel Cukier}, \bibinfo{person}{Christian K\"{a}stner}, \bibinfo{person}{Alexandros Kapravelos}, \bibinfo{person}{Dominik Wermke}, {and} \bibinfo{person}{William Enck}.} \bibinfo{year}{2025}\natexlab{}.
\newblock \showarticletitle{Research Directions in Software Supply Chain Security}.
\newblock \bibinfo{journal}{\emph{ACM Trans. Softw. Eng. Methodol.}} \bibinfo{volume}{34}, \bibinfo{number}{5}, Article \bibinfo{articleno}{146} (\bibinfo{date}{May} \bibinfo{year}{2025}), \bibinfo{numpages}{38}~pages.
\newblock
\showISSN{1049-331X}
\href{https://doi.org/10.1145/3714464}{doi:\nolinkurl{10.1145/3714464}}


\end{thebibliography}

\end{document}